\documentclass[12pt]{iopart}
\usepackage{bm}
\usepackage{epsfig}

\newcommand{\pb}{p^\mathrm{B}}
\newcommand{\pT}{p^\mathrm{T}}
\newcommand{\Gb}{G^\mathrm{B}}
\newcommand{\Gt}{G^\mathrm{T}}

\begin{document}

\title{Duality between random trap and barrier models}
\author{Robert L. Jack}
\address{Department of Chemistry, University of California at Berkeley,
Berkeley, CA 94720, USA}
\author{Peter Sollich}
\address{King's College London, Department of Mathematics, London WC2R
2LS, U.K.}

\begin{abstract}
We discuss the physical consequences of a duality between 
two models with quenched disorder, in which particles propagate
in one dimension among random traps or across random barriers. We
derive an exact relation between their diffusion fronts at
fixed disorder, and deduce from this that
their disorder-averaged
diffusion fronts are exactly equal.  We use effective dynamics schemes
to isolate the different physical processes by which particles 
propagate in the models and discuss how the duality arises
from a correspondence between the rates for these different
processes.
\end{abstract}


\section{Introduction}

How do classical particles in heterogeneous environments propagate?
This question arises in many contexts, from
glass-forming liquids and colloids~\cite{glass_expt,glass_thy}, 
to biomolecules moving in the crowded
environment of the cell~\cite{bio}, to electrical properties 
of disordered materials~\cite{Bern79}.  
Diffusion is the most familiar mode of propagation, but
if the heterogeneity involves a sufficiently broad
distribution of time scales then slower motion (subdiffusion)
is possible: for some reviews of this
well-studied field, see \cite{Alex81,HbA87,BouGeo90,MetKla00}.

Important experimental evidence of subdiffusive motion
in disordered environments was obtained a long time ago, through
the electrical conductivity of hollandite (a quasi one-dimensional ionic
conductor).  Here, the charge-carriers
can be modelled by classical particles on one-dimensional chains:
their anomalous low frequency conductance was attributed
to subdiffusive motion, and modelled by a simple one-dimensional
disordered model~\cite{Bern79}.
More recently, subdiffusive propagation has been
important when considering dynamically heterogeneous
relaxation in glass-formers.  There is considerable
evidence (for example, \cite{glass_expt,glass_thy,Qiu03,Berthier-expt-long}) 
that while time scales grow very fast with
decreasing temperature, the length
scales associated with dynamical heterogeneity increase rather 
slowly.  In the picture of the glass transition based on
dynamical facilitation~\cite{Sherrington,GC-pnas,RitSol}, 
this implies a subdiffusive
propagation of mobility through the system~\cite{subdiff-note}.
Several possible mechanisms have been proposed for
such an effect, such as directional kinetic constraints~\cite{GC-pnas},
or a strong dependence of the 
dynamics on the local free-volume~\cite{BBL}.  Thus,
understanding how subdiffusive motion arises in simple models
remains an important challenge.

In this article, we discuss the physical processes 
that lead to subdiffusive propagation in two very simple one-dimensional
models.  They are directly relevant for materials such as 
hollandite~\cite{Bern79}, and for the
models of glassy behaviour discussed in~\cite{BBL}, but
they have also been related to the motion of defects
in disordered magnets, to disordered elastic chains
and to networks of resistors and 
capacitors (see \cite{Alex81,BouGeo90} for reviews).
In the first model, diffusive motion is frustrated 
by the presence of sites where the particle
gets trapped for a long time; in the second, there are 
barriers between sites across which motion is very slow.
These models are related by a duality relation between
their master operators.  This relation has been 
noted before~\cite{Alex81,BouGeo90,Dyson53}, 
but with limited discussion
of its physical content.  Here, we revisit the duality
between the two models, focussing on their diffusion
fronts, and derive exact relations between these which do not appear to
have been noticed previously. We connect the evolution of the
diffusion fronts with the events in which
the particle escapes from deep traps or surmounts large barriers.
We elucidate the physical processes underlying the duality,
and use it to motivate an effective dynamics for the model,
following le Doussal, Monthus and Fisher~\cite{DMF99}.

We define our models in Section~\ref{sec:model}
and discuss the duality relation in Section~\ref{sec:dual}.  We
cast the duality relation for fixed disorder as a local relation 
between the propagators of our models, and show 
that it leads to equal disorder-averaged propagators in the two systems. 
In Section~\ref{sec:eff}, we describe the physical processes
by which the models evolve, and use them to define effective dynamics
schemes. The extent to which this
dynamics captures the real-time evolution of the diffusion front is
also discussed.
We summarise our conclusions in Section~\ref{sec:conc}.

\section{Models}
\label{sec:model}

We first define the random trap model~\cite{BouGeo90}.  Consider
a single particle hopping on a chain of sites.  
A particle on site $n$ may hop 
either to the left or right, and hops in both directions happen 
with the same (site-dependent) rate, $W_n$.  The master equation 
for the distribution of particle positions, $p_n^\mathrm{T}$, is then
\begin{equation}
(\mathrm{d}/\mathrm{d}t) \pT_n(t) 
= W_{n+1} \pT_{n+1}(t) + W_{n-1} \pT_{n-1}(t) - 
                                 2W_n \pT_n(t)
\label{equ:w_trap}
\end{equation}
The superscript `T' distinguishes quantities for the trap model from
those for the barrier model. In the latter~\cite{Alex81,BouGeo90}, 
the rates $W_n$ are associated
not with sites, but with links between them.  That is, hops from
site $n$ to $n+1$ happen with rate $W_n$, as do hops from
site $n+1$ to $n$.  In this case, the master equation is
\begin{equation}
(\mathrm{d}/\mathrm{d}t) \pb_n(t) 
  = W_{n} [\pb_{n+1}(t)-\pb_n(t)] 
                                + W_{n-1}  [\pb_{n-1}(t)-\pb_n(t)] 
\label{equ:w_barrier}
\end{equation}

In~\cite{BouGeo90}, it was noted that if the same
set of rates $W_n$ are used in both models, then 
the current in the barrier
model $j^\mathrm{B}_n=W_n(\pb_n-\pb_{n+1})$ obeys the same
differential equation as the rescaled probability
density $W_n \pT_n$ in the 
trap model.  This duality between trap and barrier models
was noted a long time ago by Dyson~\cite{Dyson53}: we refer
to it as `trap-barrier duality'. As explained below, it follows
that the eigenspectra of the two master operators
are equal for these two models: this property
is relevant for the impedances of linear chains of 
resistors and capacitors, and for the elasticity of random 
one-dimensional chains.  However, in understanding how
particles propagate in disordered media, the most natural
quantity is the propagator, or diffusion front, which we discuss
in some detail below.

It was further argued in~\cite{BouGeo90} that trap-barrier duality can be
understood by identifying the regions between large barriers in one 
model with the deep traps in the other.  While very appealing
intuitively, this argument is possibly somewhat misleading, in two
opposite ways. On the one hand, it suggests that the duality holds only after
suitable coarse-graining to large lengths and times; we will see, however,
that for the diffusion fronts it is valid on {\em all} length and time
scales. On the 
other hand, viewing a barrier model as consisting of effective traps
would suggest that, at least on large enough scales, the models will
behave effectively identically. This also is not quite correct.
For example, the barrier model has a stationary 
distribution in which the particle occupies each site with
equal probability, while the trap model exhibits aging behaviour if it
is initialised in a uniform state~\cite{MonBou96}. Quantitatively,
consider a particle that starts
from the origin at time zero, and measure its mean square displacement
between times $t'$ and $t>t'$.  Averaging over the random rates $W_n$
according to a translationally invariant distribution, we have 
(see Section~\ref{sec:dual}) that
\begin{equation}
R^2(t,t') = \overline{ \langle [x(t)-x(t')]^2 \rangle } 
\label{equ:2time_disp}
\end{equation}
depends only on the time difference $t-t'$ in the barrier model
(the angle brackets represent an average over the stochastic
dynamics, and the overbar represents the average over the
rates $W_n$).  On the other hand, in the trap model, $R^2(t,t')$ 
generically depends on both time arguments, and for the disorder
distributions we will consider it never
reaches a stationary regime in the thermodynamic limit.

Bearing this and other differences in mind, we now
use the relationship between (\ref{equ:w_trap}) and 
(\ref{equ:w_barrier}) to investigate the
extent to which the barrier model can be modelled by a set of
effective traps.  We establish
relationships between the propagators (diffusion fronts) of the
two models, and use this analysis to develop a physical picture
of propagation in these systems.
We also show that the duality implies not just that the
long-time behaviour of these models are the same, but
that their disorder averaged propagators are identical at all times.  
We contrast the extent to which the models are similar with
differences between them, including their aging behaviour
and the properties of the typical particle trajectories.

\section{Duality}
\label{sec:dual}

It is convenient to define the master operator for the
barrier model, $\hat{W}^\mathrm{B}$,
in terms of its matrix elements $W^\mathrm{B}_{nm}$, by writing
(\ref{equ:w_barrier}) as 
$(\mathrm{d}/\mathrm{d}t)\pb_n(t)=-\sum_m W^\mathrm{B}_{nm} \pb_m(t)$.
We specify periodic boundaries on a chain of length $L$, so
this operator has $L$ eigenvectors $q^{\rm B}_n$. The eigenvector with
zero eigenvalue gives the steady state $q^\mathrm{B,eq}_n=1/L$. The
dynamics of the model obeys detailed balance 
with respect to this trivial distribution, as is clear from
the symmetry $W^\mathrm{B}_{nm}=W^\mathrm{B}_{mn}$. For the trap model
we call the master operator $\hat{W}^\mathrm{T}$; it too obeys
detailed balance but with respect to the non-uniform steady state
$q^\mathrm{T,eq}_n \propto W_n^{-1}$.

The expression of trap-barrier duality in terms of propagators can be
derived from the simple fact that, if $p^\mathrm{B}_n(t)$ is a
solution of the barrier model master equation
(\ref{equ:w_barrier}) then $p^\mathrm{B}_n(t)-p^\mathrm{B}_{n+1}(t)$
is a solution of the trap model equation (\ref{equ:w_trap}). Applying
this to a barrier model eigensolution $p^\mathrm{B}_n(t)=q^\mathrm{B}_n
\exp(-\lambda t)$, it follows that
$q^\mathrm{T}_n=q^\mathrm{B}_n-q^\mathrm{B}_{n+1}$ is an eigenvector for
the trap model master operator with the same eigenvalue
$\lambda$. This is a one-to-one relation between the eigenvectors with
nonzero eigenvalues of the two master operators, and so their spectra
are identical as claimed in Section~\ref{sec:model}. (Inverting the
differencing relation to get 
from $q^\mathrm{T}$ back to $q^\mathrm{B}$ in principle gives an
undetermined constant, but this is fixed by the requirement that
nonzero eigenvectors obey $\sum_n q^\mathrm{B}_n=0$.) The exception is
the steady state, where differencing the barrier model steady state
$q^\mathrm{B,eq}=1/L$ gives a vanishing result rather than the trap
model steady state.

\begin{figure}
\epsfig{file=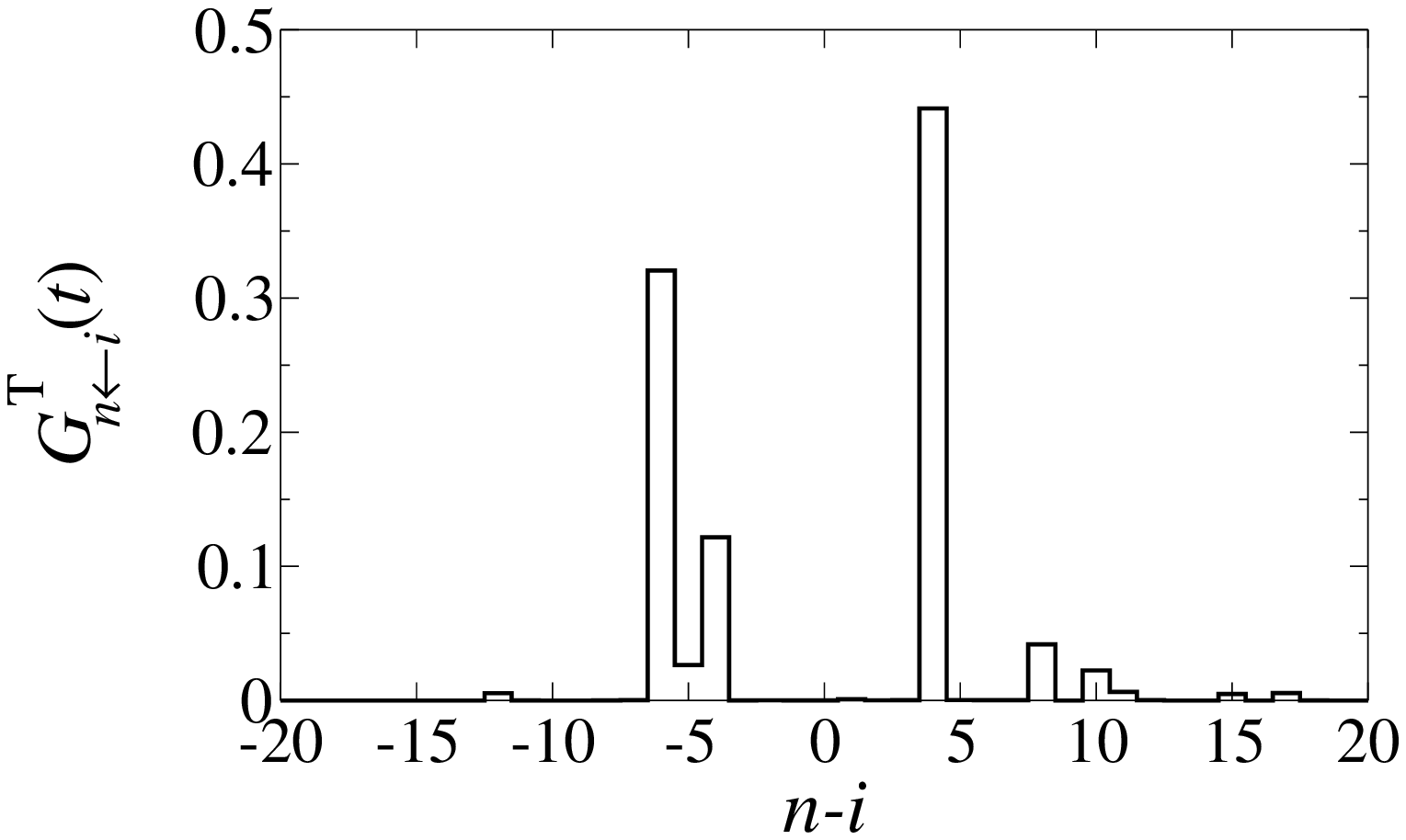,width=7.5cm}
\epsfig{file=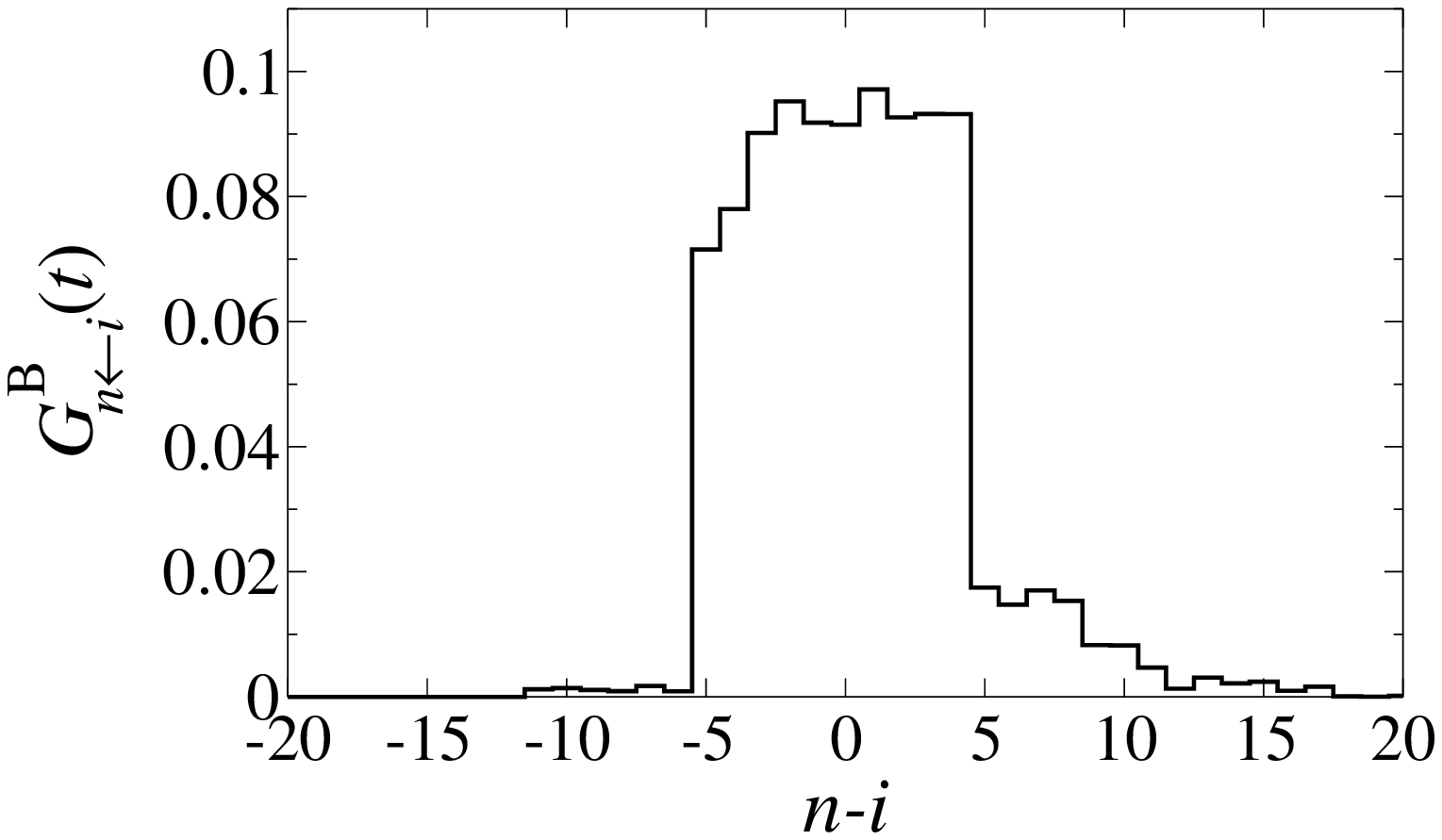,width=7.5cm}
\par
\epsfig{file=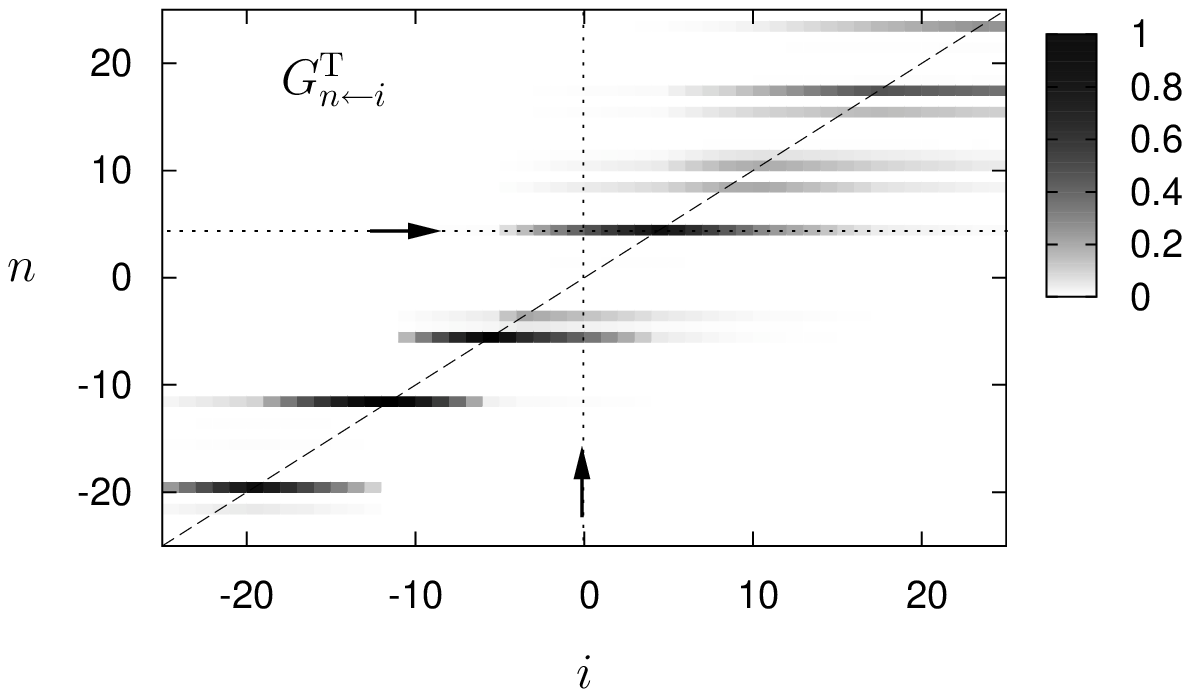,width=8cm}
\epsfig{file=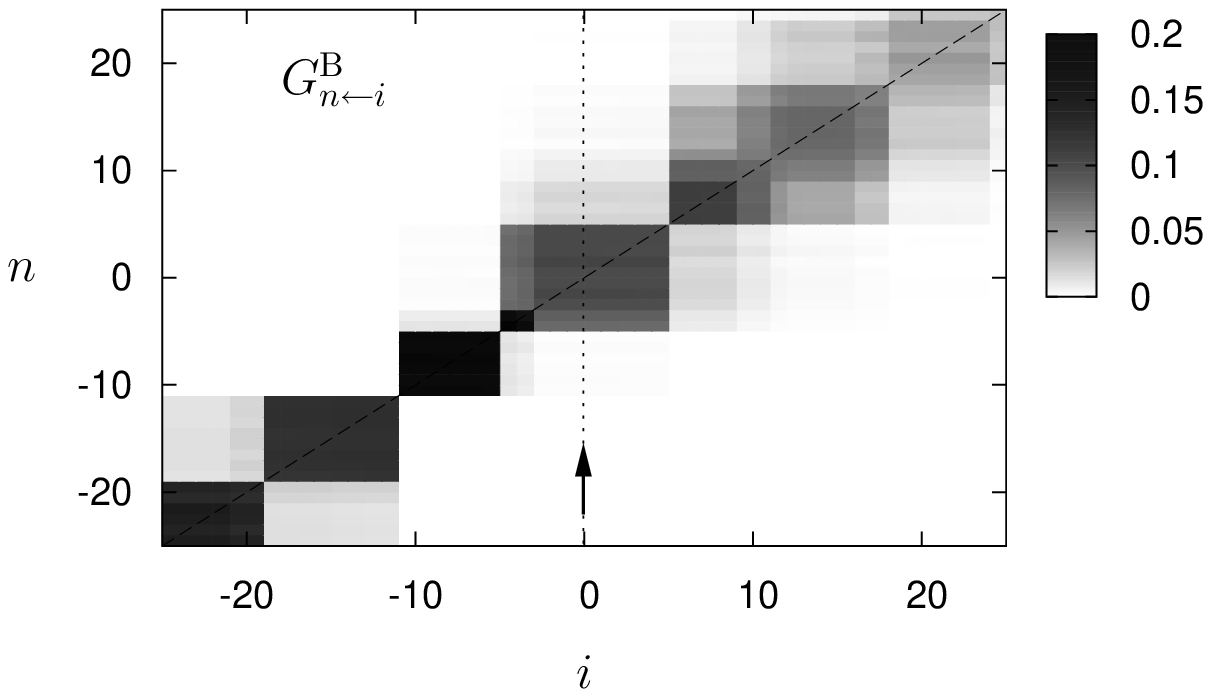,width=8cm}
\par
\hspace{3.75cm}\epsfig{file=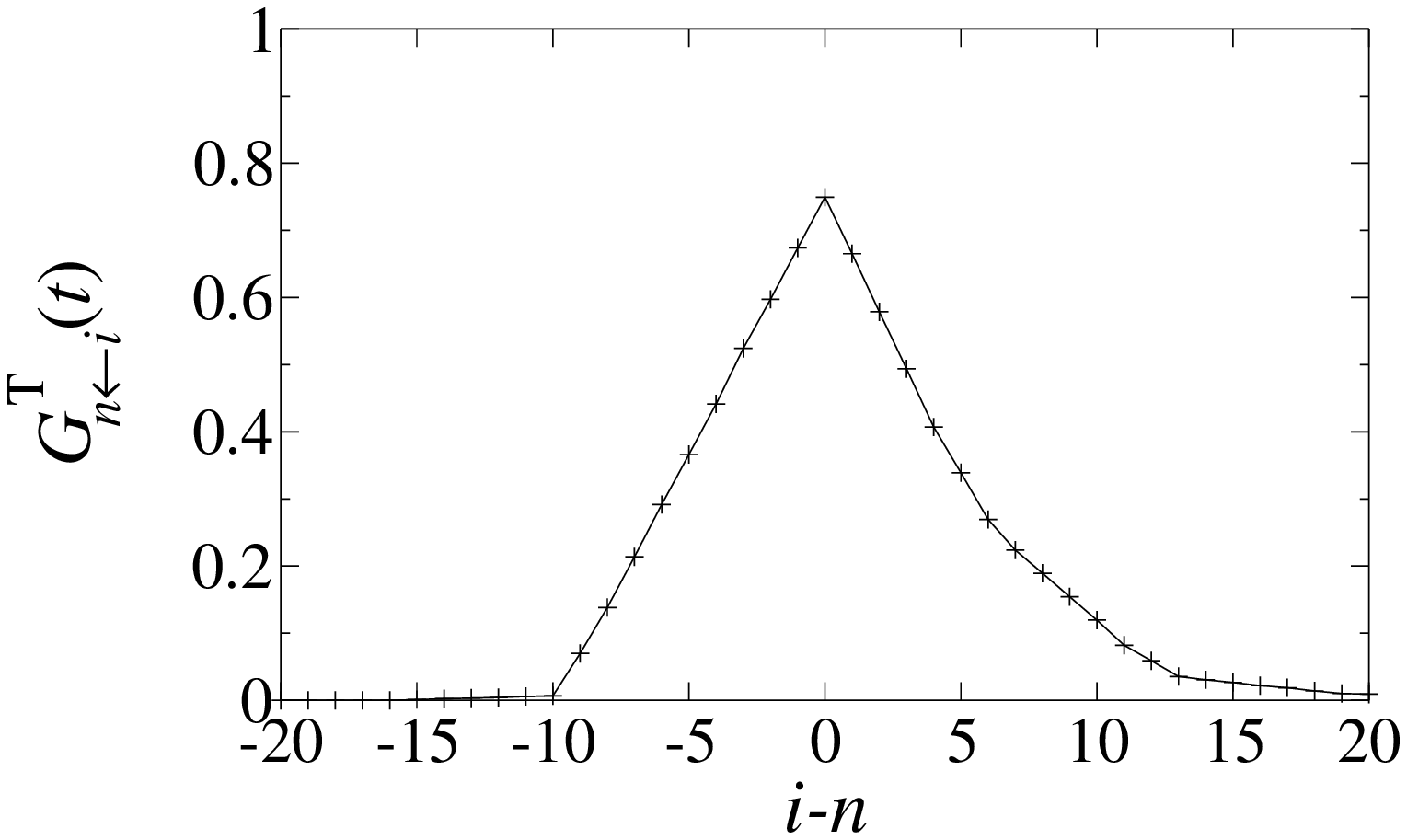,width=7.5cm}
\caption{
(Top)~Sample propagators, $\Gt_{n\leftarrow i}$
and $\Gb_{n\leftarrow i}$, as a function of final
site $n$, for trap and barrier models at $\mu=3$.
The set of disordered rates is the same in each case, and
we take $i=0$ and $t=2^{14}$.  Small rates lead to sites with peaks in the trap
model propagator, and to large changes between adjacent sites in the 
barrier model.
(Middle)~Propagator maps for the same systems: the propagators at
fixed time are plotted as functions of both initial and final
positions.  The dashed lines indicate $n=i$ and the dotted lines
with arrows indicate the slices for which we plot data in the other panels
of this figure.
In the trap model, the probability is concentrated
on final sites with small escape rates.  In the barrier model, it is
delocalised 
across `effective traps', separated by barriers with small crossing rates.  The
duality relation (\ref{equ:local_dual}) states that the gradient of
the trap model propagator with respect to $i$ is equal to the
negative gradient
of the barrier model propagator with respect to $n$.  The relevant
gradient in the barrier model is localised on the upper and lower edges of the 
effective traps, and these are the final sites on which the probability
is localised in the trap model.  
(Bottom)~We plot the trap
model propagator as a function of initial site $i$,
for a fixed final site $n=4$.  This illustrates its gradient
with respect to $i$.  
}
\label{fig:gfs}
\end{figure}

The propagator or diffusion front of the barrier model,
$\Gb_{n\leftarrow i}(t)$, is defined as the solution of
(\ref{equ:w_barrier}) with initial condition $\pb_n(0)=\delta_{n,i}$.
This can also be obtained from matrix elements of the time evolution
operator: $\Gb_{n\leftarrow i}(t)=[\exp(-\hat{W}^\mathrm{B}t)]_{ni}$.
The symmetry of the barrier master operator $\hat{W}^\mathrm{B}$ then
implies that also the propagator is symmetric under interchange of
arrival and departure sites
\begin{equation}
\Gb_{n\leftarrow i}(t)=\Gb_{i\leftarrow n}(t)
\label{equ:barrier_symmetry}
\end{equation}
as required for a system obeying detailed balance with
respect to a uniform distribution.

We now turn to the trap-barrier duality relation for the propagators.
Since $\pb_n(t)=\Gb_{n\leftarrow i}(t)$
is the solution to (\ref{equ:w_barrier})
with initial condition $\pb_n(0)=\delta_{n,i}$, 
it follows that
$\pT_n(t)=\Gb_{n\leftarrow i}(t)-\Gb_{n+1\leftarrow i}(t)$
is a solution of (\ref{equ:w_trap}), with initial condition 
$\pT_n(0)=\delta_{n,i}-\delta_{n+1,i}=\delta_{n,i}-\delta_{n,i-1}$.
Defining the propagator 
for the trap model $\Gt_{n\leftarrow i}(t)$ as the solution
to (\ref{equ:w_trap}) with initial condition $\pT_n(0)=\delta_{n,i}$, it
follows (by linearity of the master equation) that 
\begin{equation}
\Gb_{n\leftarrow i}(t)-\Gb_{n+1\leftarrow i}(t)
 = 
\Gt_{n\leftarrow i}(t)-\Gt_{n\leftarrow i-1}(t)
\label{equ:local_dual}
\end{equation}
This identity, which holds for all realisations of the
disorder $\{W_n\}$, is our desired exact statement of the trap-barrier
duality relation, in terms of the propagators of the two
models. It can be used to express the trap model propagator in
terms of the barrier model one, and vice versa.  For example,
since the diffusion front vanishes at large distances, we have
\begin{equation}
\Gb_{n\leftarrow i}(t)=\sum_{j=n}^{L/2} [\Gb_{j\leftarrow i}(t)
-\Gb_{j+1\leftarrow i}(t)]=\sum_{j=n}^{L/2} [\Gt_{j\leftarrow i}(t)
-\Gt_{j\leftarrow i-1}(t)].
\label{equ:gbar_inf}
\end{equation}
(We require that the chain length $L$ is 
large enough that $\Gb_{j\leftarrow i}$ does indeed vanish at large $j$.
That is, $L$ should be taken to infinity
before taking any limit of large time.) A corresponding relation
can be used to express $\Gt$ in terms of $\Gb$. We show typical propagators 
in Fig.~\ref{fig:gfs}, and give a geometrical interpretation
of~(\ref{equ:local_dual}):
the difference of the barrier model propagator taken between
neighbouring {\em arrival} sites equals the negative difference of the trap
model propagator between neighbouring {\em departure} sites.

In making the plots in Fig.~\ref{fig:gfs} we considered the most common case
where the $W_n$ are independently and identically distributed (i.i.d.), with
a distribution 
\begin{equation}
P(W) \propto W^{(1/\mu)-1}, \qquad 0<W<1
\label{equ:P_W}
\end{equation} 
With this choice, the long time behaviour of a particle is diffusive
for $\mu<1$ and subdiffusive for $\mu>1$.  More precisely, if a particle
begins on a randomly chosen site, its displacement at large times
scales as~\cite{BouGeo90,Machta85}
\begin{equation}
\langle x^2(t) \rangle \sim t^{2/z}, \qquad z=\max(2,1+\mu)
\label{equ:scaling}
\end{equation}
with logarithmic corrections for $\mu=1$.

It is sometimes useful to interpret the rates $W_n$ as arising
from activation 
energies $E_n$ associated with crossing a barrier or leaving a trap.
We then write
$W_n=\exp(-E_n/T)$, where $T$ is the temperature and we have
set Boltzmann's constant equal to unity.  In this representation,
assuming the distribution of the
activation energies is $P(E)=\exp(-E)$ with $E>0$,
we can identify $\mu$ with $1/T$.

We now consider averaging the propagators $\Gb$ and $\Gt$
over realisations of the disorder.
For any distribution of the disorder which is translationally invariant
(which includes all choices in which the $W_n$ are i.i.d.),
the disorder-averaged propagators
depend only on the difference $m=n-i$.
Using a bar to denote the disorder average as before, we thus have
from the duality relation (\ref{equ:local_dual})
\begin{equation}
\overline{G}^\mathrm{B}_{m}(t)-\overline{G}^\mathrm{B}_{m+1}(t)
 = 
\overline{G}^\mathrm{T}_{m}(t)-\overline{G}^\mathrm{T}_{m+1}(t).
\end{equation}
Since both propagators must vanish as $|m|\to\infty$,
it follows that
\begin{equation}
\overline{G}^\mathrm{B}_{m}(t)=\overline{G}^\mathrm{T}_{m}(t)
\label{equ:ave_dual}
\end{equation}

This exact coincidence of the disorder averaged diffusion fronts is a
surprising result. For a given realisation of the disorder, the
propagators for the two models are very different, as shown in
Fig.~\ref{fig:gfs}. For other quantities the difference is even more
striking. Consider for example the quantity $R^2(t,t')$ defined in
(\ref{equ:2time_disp}) above. Since all propagators decay to zero at
large distances, we can replace the disorder average by an average
over initial position $i$, so that $R^2(t,t')=(1/L)\sum_{k,n}
(x_n-x_k)^2 G_{n\leftarrow k}(t-t') \sum_i G_{k\leftarrow i}(t')$,
where $x_n$ is the position of site $n$. But
now the symmetry (\ref{equ:barrier_symmetry}) of the barrier model
propagator together with conservation of probability implies $\sum_i
G_{k\leftarrow i}(t')= \sum_i G_{i\leftarrow k}(t')=1$.  Thus
$R^2(t,t')$ depends on $t$ and $t'$ only through the propagator $
G_{n\leftarrow m}(t-t')$, and therefore only through the time
difference $t-t'$. The argument evidently generalizes to the entire
(disorder-averaged) distribution of $x(t)-x(t')$.  This
time translation invariance arises because
a particle moving among barriers is equally likely to be on any site,
whether or not it is bordered by a high barrier. In the trap model, on
the other hand, the particle spends (for $\mu>1$) most of its time in
deep traps whose escape rate decreases with the age $t'$ as $W\sim
t'{}^{(1/z)-1}$~\cite{BouGeo90}. 
To arrive at the asymptotic scaling 
of $R^2(t,t')$ in the trap model, we take $t>t'$, and 
note that moves to left and right
are always equally likely, so that 
$\langle [x(t)-x(t')]x(t')\rangle=0$ (for any realisation
of the disorder, as
long as the system size $L$ is taken to infinity before
any limit of large time~\cite{parity-proof}).  Thus, 
$R^2(t,t')=\overline{\langle x^2(t)\rangle} - \overline{\langle
x^2(t')\rangle}$, which 
for large $t$ and $t'$ scales as
$t^{2/z}-t'^{2/z}$.  Clearly, time translational
invariance is obtained only in the diffusive case ($z=2$ or $\mu<1$).

The difference between the two models in the subdiffusive regime
$\mu>1$ can also be seen at the level of typical trajectories. For
example, let $N$ 
be the mean number of
hops made by a particle up to time $t$.  Time
translation invariance in the barrier model implies that $N\sim t$.
To get the scaling (\ref{equ:scaling}), the displacement must
then grow with the number of hops as $|x|\sim N^{1/z}$, i.e.\ more
slowly than for simple diffusion.  On the other hand,
in the trap model, 
the typical number of hops associated with trajectories of 
displacement $|x|$ scales as in simple diffusion: $N\sim|x|^2$, because
escapes from any trap occur with equal probability to the left and the
right. This then implies a number of hops growing
only sublinearly in time, $N\sim t^{2/z}$~\cite{BouGeo90}.

To summarize thus far, the trap and barrier models have on average equal
diffusion fronts, but many other physical properties differ.
Motivated by this surprising observation, 
we now discuss the duality relation at fixed disorder in more detail,
in order to understand how the equality of the disorder-averaged
diffusion fronts arises.

\section{Effective dynamics}
\label{sec:eff}

In Fig.~\ref{fig:gfs} we showed the propagators for trap and barrier models,
for a single realisation of the disorder (with $\mu=3$) and at a
given time.
%
To understand how the propagators evolve in time
we exploit an `effective dynamics picture'
in the spirit of~\cite{DMF99}.  
A very similar scheme was applied to the trap
model in~\cite{Monthus03,Monthus04}.  The
effective dynamics is based on an assumption of well-separated
time scales, which is valid in the limit of large $\mu$ taken at fixed
energies $E_n$. We first present the scheme for the barrier model. 

\subsection{Barrier model}

\begin{figure}
\epsfig{file=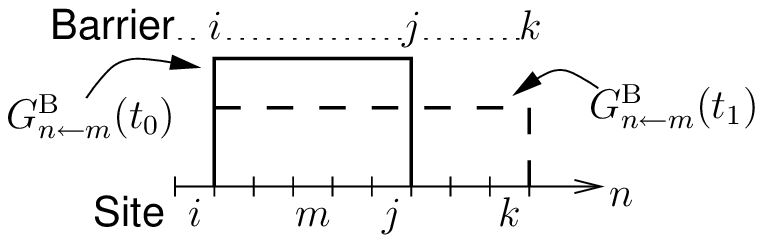,width=7.5cm} \hspace{12pt}
\epsfig{file=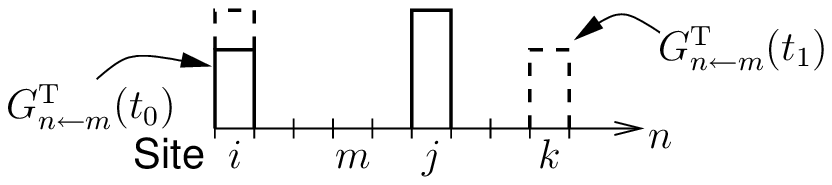,width=7.5cm}
\caption{(Left) Sketch illustrating a single step of the effective
dynamics for the barrier model.  Barriers $i$, $j$ and $k$ are all relevant
at time $t_0$, so $\Gb_{n\leftarrow m}(t_0)$ is localised between barriers
$i$ and $j$.  At the later time $t_1$, barrier $j$ has become irrelevant
($K_j t_1>1$).  (Right) A sketch of the corresponding
behaviour for the trap model.  The propagator is localised
on sites $i$ and $j$ at the initial time $t_0$ and on
sites $i$ and $k$ at the later time $t_1$.  The value of the 
propagator on these sites is in inverse proportion
to their distance from the initial site $m$.
}
\label{fig:barr_sketch}
\end{figure}

To define the effective dynamics  in this 
model, we separate those barriers
which are relevant on a time scale $t$ from those
which are irrelevant.  Large barriers (those with small 
transmission rates $W$)
tend to be relevant, because they limit the motion of
the particle.  Once we have identified a set of relevant
barriers, we assume that particles move rapidly between
them, but never cross them.  Thus, for an initial site $m$,
we have
\begin{equation}
\Gb_{n\leftarrow m}(t) = \left\{ \begin{array}{ll}
  (j-i)^{-1}, \qquad & i< n\leq j \\ 0, & \hbox{otherwise}
  \end{array}\right.
\label{equ:eff_prop_barr}
\end{equation}
where $i$ is the index of the nearest relevant barrier to the left of 
site $m$, and $j$ the index of the nearest relevant barrier to its 
right (see Fig.~\ref{fig:barr_sketch}). 
These two barriers define the `effective trap' within
which the particle is localised.
In the propagator map representation of Fig.~\ref{fig:gfs}, the effective
dynamics models the propagator as a series of non-overlapping
blocks.  As the set of relevant barriers evolves in time, 
the map evolves by instantaneous events in which two
adjacent blocks coalesce into a single larger one.

To obtain the time scale on which barriers cease to be relevant,
suppose that the two barriers delimiting
the effective trap have well-separated transmission rates, 
$W_j\gg W_i$.  On time scales much smaller than $W_i^{-1}$,
the diffusion front evolves in time only by transmission
through barrier $j$.  The diffusion front
increases within a neighbouring effective trap, which is formed by
barrier $j$, and the nearest relevant barrier
to its right (let the index of this barrier be $k$, as in
Fig.~\ref{fig:barr_sketch}).
We now further assume that the motion between barriers $j$ and $k$
is fast compared to transmission through them.  In that
case, we can approximate the propagator by
\begin{equation}
\Gb_{n\leftarrow m}(t) = 
  \left\{ \begin{array}{ll} \rho_0(t),& \qquad i<n\leq j \\
               \rho_1(t),& \qquad j<n\leq k \end{array}\right.
\label{equ:g_rho_t}
\end{equation}
where $\rho_0$ and $\rho_1$ are probability densities within the
two effective traps (to the left and right of barrier $j$: see
Fig.~\ref{fig:barr_sketch}).  
The probabilities in these traps are
$\rho_0 l_0$, and $\rho_1 l_1$, where $l_0=j-i$ and $l_1=k-j$ 
are the sizes of the effective traps.  These probabilities evolve as
$\partial_t [l_0 \rho_0(t)] = -\partial_t [l_1\rho_1(t)] = 
W_j[\rho_1(t)-\rho_0(t)]$.  Hence,
\begin{equation}
\partial_t [\rho_0(t)-\rho_1(t)] = -K_j [\rho_0(t)-\rho_1(t)],
\end{equation}
and we identify
\begin{equation}K_j=W_j\left(\frac{1}{l_0}+\frac{1}{l_1}\right)
\label{equ:kj}\end{equation}
as the rate for equilibration between these two effective traps.

Still working within our assumptions of well-separated time scales,
we can use this procedure to define the time evolution of the set
of relevant barriers.  At small times, all barriers are relevant.  For
a given set of relevant barriers, we characterise barrier $j$ by its value 
of $K_j$, where the lengths $l_0$ and $l_1$ are the distances to
the nearest relevant barriers to the left and right of barrier $j$.
(When a relevant barrier is removed, a new effective
trap is formed, and the values of $K_j$ on
the adjacent barriers change according to the width of the new trap.)
Thus the effective dynamics evolves in time by successive removal 
of the barrier with the largest value of $K_j$.  
To obtain the set of relevant barriers at time $t$,
this procedure is iterated until all relevant 
barriers have $K_j t<1$.  
The propagator is then constructed
using (\ref{equ:eff_prop_barr}), as described above.

This approach is similar to that of~\cite{DMF99}, differing only in that the
sizes of the effective traps enter the transition rates.  This is necessary
for this model, since these factors affect the scaling of the diffusion front
(see Section~\ref{subsec:ave}).  
We also note that the rate $K_j$ depends symmetrically on the
width of the effective traps to the left and the right of the barrier. 
This ensures that the effective dynamics for the diffusion
front preserves its symmetry (\ref{equ:barrier_symmetry}).

We now revisit our assumptions regarding
well-separated time scales.  These assumptions were (i) that the two
barriers delimiting each effective trap have well-separated
rates, allowing us to treat transmission over them independently, (ii) the rates associated
with successive stages of the effective dynamics are well-separated,
and so the time taken for a given barrier to become irrelevant
is well-approximated by $K_j^{-1}$, and
(iii) equilibration within each of two neighbouring effective traps is faster
than transitions between them [so that (\ref{equ:g_rho_t}) can
be applied]. Writing $W_n=\exp(-E_n/T)$ as before, with $T=1/\mu$,
we recall that the distribution of energies is $P(E_n)=\exp(-E_n)$.
For the first assumption, the time scales for crossing
adjacent barriers are well-separated if the difference
between their two energies is much greater than $T$. In the limit
of low temperature (or large $\mu$) the probability of finding
two adjacent barriers with rates within $T$ of each other vanishes, so this
assumption holds.  For the second
assumption, the same argument implies that the rate for any stage of the 
dynamics is well-separated from the (larger) 
rate of the previous stage.
Finally, for the third assumption, the rate for
equilibration within an effective trap is of the order of the
rate of a previous stage of the effective dynamics, which
is well-separated from the rate of the current stage as per the
second assumption.

\begin{figure}
\epsfig{file=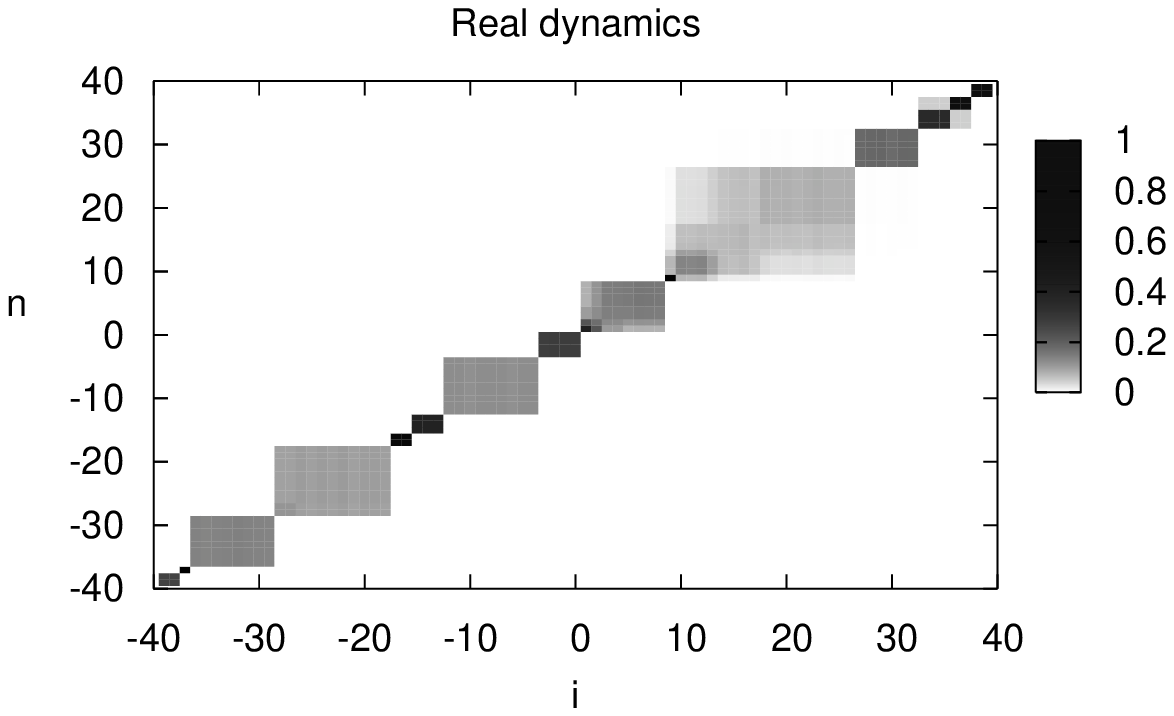,width=8cm}
\epsfig{file=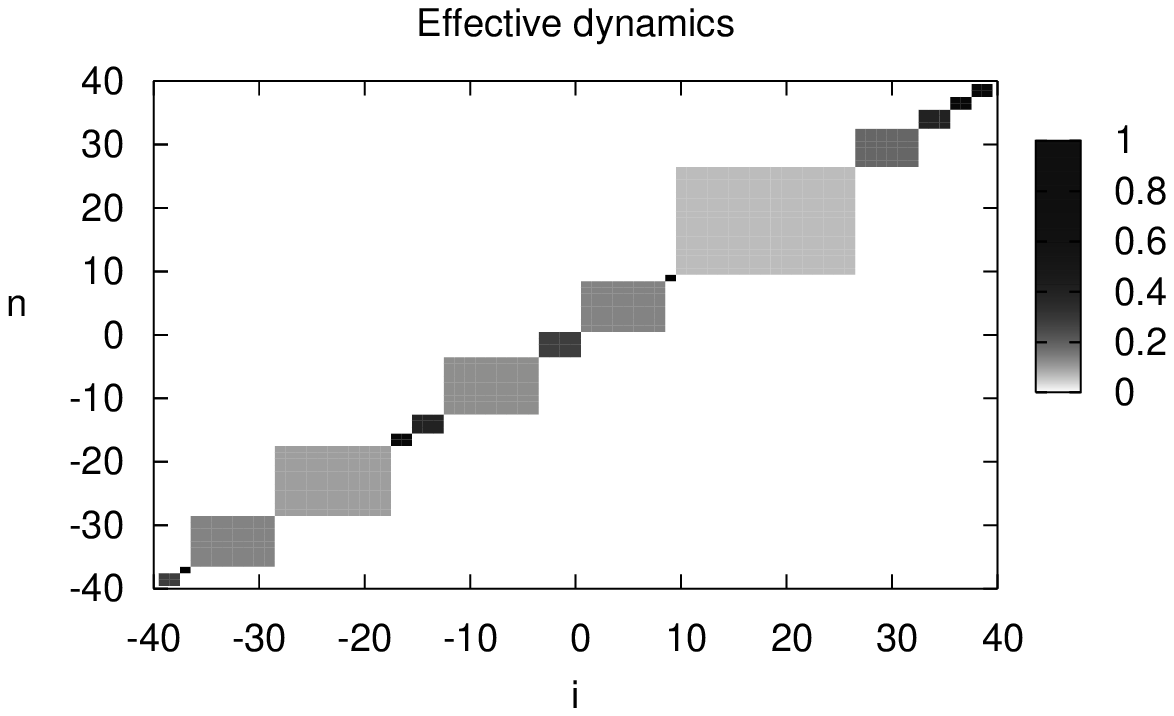,width=8cm}
\caption{(Left) Propagator map for
the barrier model at $\mu=10$ and $t=2^{29}$.  This
was calculated by histogramming the corresponding
trap model propagator and using~(\ref{equ:local_dual}).
(Right) Propagator map for the effective
dynamics, using the same realisation of the disorder
and the same final time.  The agreement is quite good,
although the real propagator has some fine structure
not captured by the effective dynamics.}
\label{fig:gb_eff}
\end{figure}

In Fig.~\ref{fig:gb_eff}, we compare the propagator for a given
realisation of the disorder with
the prediction of the effective dynamics, at
$\mu=10$.  The agreement is quite good, justifying our
assumptions of well-separated time scales at this value of $\mu$.
As $\mu$ is decreased from this large value towards unity, these assumptions
break down.  The result is that the propagator cannot be well-approximated
by non-overlapping blocks in the density map representation 
(recall the structure
of the propagator at $\mu=3$, shown in
Fig.~\ref{fig:gfs}).  We discuss the agreement between
effective and real dynamics in more detail in Section~\ref{subsec:ave},
at the level of the 
disorder-averaged diffusion front. First, though, we construct
an effective dynamics for the trap model, related to the effective 
dynamics of the barrier model by the duality 
relation~(\ref{equ:local_dual}).

\subsection{Trap model}

Since we have the duality relation (\ref{equ:local_dual}), the effective
dynamics that we have described for the propagator of the barrier model has 
a corresponding effective dynamics in the trap model.  
We begin by describing this effective dynamics in terms of physical
processes in the trap model, and then we show that it is indeed 
the dual of that of the barrier model.

We assume that for a given initial site $m$, the propagator
is concentrated on two `relevant' traps on sites
$i$ and $j$, with $i<j$.  As before, if time scales are well-separated then
we can consider these two sites separately.  If $W_i\ll W_j$,
the relevant
physical processes are those in which the particle escapes from
trap $j$ and falls into another relevant trap: either
the one on site $i$, or one located on a site $k>j$  (see
Fig.~\ref{fig:barr_sketch}).
The particle hops from site $j$ to $j+1$ with rate $W_j$.  However,
the probability of it reaching site $k$ before being reabsorbed on
site $j$ is $1/(k-j)$.  
(Particles always hop left and right with equal 
probability, so this combinatorial factor is the same as for 
simple diffusion.) Thus, the rates for motion from site $j$
to sites $i$ and $k$ are $W_j/(j-i)$ and $W_j/(k-j)$ respectively.
We therefore identify the total rate for moving from site $j$ 
to either neighbouring trap as $K_j$, as defined in (\ref{equ:kj}).  
This is the time scale on which the trap
on site $j$ becomes irrelevant.  Hence, the effective dynamics
for the set of relevant rates in the trap model
is the same as the effective dynamics for those in the barrier model.

It remains to find the value of the propagator 
on these relevant sites. Starting
from a single trap on site $m$, with $i<m<j$, the rates for arriving
on sites $i$ and $j$ are in inverse proportion to their
relative distances from $m$. Therefore, if at time $t_0$ trap
$m$ has become irrelevant then 
$\Gt_{i\leftarrow m}(t_0) = (j-m)/(j-i)$ and
$\Gt_{j\leftarrow m}(t_0)=(m-i)/(j-i)$.  In the next step of the process,
site $j$ becomes irrelevant, by time $t_1$ say, and the probability on
that site is redistributed onto sites $k$ and $i$, with probabilities
inversely proportional to their distances from site $j$.
Thus, we can relate the 
propagator at $t_1$ to that at $t_0$ by
$\Gt_{i\leftarrow m}(t_1) = \Gt_{i\leftarrow m}(t_0)
+ \frac{k-j}{k-i}G_{j\leftarrow m}(t_0) = \frac{k-m}{k-i}$ and 
$\Gt_{k\leftarrow m}(t_1) = \frac{j-i}{k-i}\Gt_{j\leftarrow m}(t_0)
=\frac{m-i}{k-i}$.  
The relative probabilities of the two relevant sites at time $t_1$
are thus in inverse proportion to their distances from the initial
site $m$, just as they were at time $t_0$.
Hence, this property is maintained as successive barriers
become irrelevant.

These arguments motivate us to define our effective dynamics 
for the diffusion front as:
\begin{equation}
\Gt_{n\leftarrow m}= \frac{j-m}{j-i} \delta_{n,i} + \frac{m-i}{j-i} \delta_{n,j}
\label{equ:eff_prop_trap}
\end{equation}
where $i$ and $j$ are the nearest relevant traps to the 
left and right of the initial site $m$, and the
set of relevant traps is determined by successively
removing the trap with the smallest value of $K_j$.  
A similar result was derived in~\cite{Monthus03}, 
although relevant traps were identified by their values 
of $W_j$ in that work.  These choices seem to be equivalent in the 
limit of large $\mu$ in which the effective dynamics schemes are valid.  
Returning to our analysis,
it is then trivial to check that (\ref{equ:eff_prop_trap})
and (\ref{equ:eff_prop_barr}) are consistent with the duality
relation (\ref{equ:local_dual}), and that they are therefore
dual to each other.  The comparison between effective  and
real dynamics for the {\em barrier} model was shown in 
Fig.~\ref{fig:gb_eff}.  Since both real and effective
dynamics obey the duality relation, the agreement
is similarly good for the trap model.

Thus, we have identified the physical processes leading to motion in the
trap model, and used these processes to motivate an effective dynamics
for the diffusion front.  This effective dynamics is 
 related by the trap-barrier duality of (\ref{equ:local_dual})
to that defined in the previous section for the barrier model.
We trace this symmetry to the equality between the rates $K_j$
for the (different) physical processes that occur in the two models.

\subsection{Disorder-averaged diffusion fronts}
\label{subsec:ave}

Turning to disorder-averaged properties of the effective dynamics,
the scaling (\ref{equ:scaling}) for $\mu>1$ arises directly from
the form of the rate $K_j$ given in (\ref{equ:kj}).  To demonstrate
this, we use the language of the barrier model.  Let
the typical distance between relevant barriers at time $t$
be $\langle \ell(t) \rangle$.  (Unlike distributions of times,
all moments of the distribution of lengths are finite, 
so scaling arguments based on typical widths are valid.)
Since relevant barriers have $Kt<1$,
the rates $W$ associated with these barriers typically
satisfy $Wt<\langle \ell(t)\rangle$.  For consistency, the typical
distance between these barriers must scale as $\langle\ell(t)\rangle$,
so we have
\begin{equation}
\langle \ell(t)\rangle^{-1} \sim \int_0^{\langle\ell(t)\rangle/t} 
  P(W) \mathrm{d}W
\end{equation}
where $P(W)$ was given in (\ref{equ:P_W}).
This leads to $t\sim\ell(t)^{1+\mu}$.  All length scales in the effective
dynamics scale with $\ell(t)$, so this dynamics leads to $z=1+\mu$,
giving the correct scaling of the diffusion front for all $\mu>1$ 
[recall (\ref{equ:scaling})].

Following the reasoning in~\cite{DMF99}, scaling arguments can also be
used to express the disorder-averaged diffusion front in terms of the
fraction of effective traps of length $\ell$ at time $t$.
%
%
In the scenario studied in~\cite{DMF99,Monthus03},
where the widths of all effective
traps and the rates of the relevant barriers are all
independent, this allows one to make significant progress.
By contrast, in our effective
dynamics the relevance of barriers is determined by $K_j$,
and this introduces correlations between the rates and
the widths of the effective traps. We have therefore not been able
to determine the required distribution of trap lengths $\ell$
analytically, and instead return to numerical simulations.

\begin{figure}
\hspace{4cm}
\epsfig{file=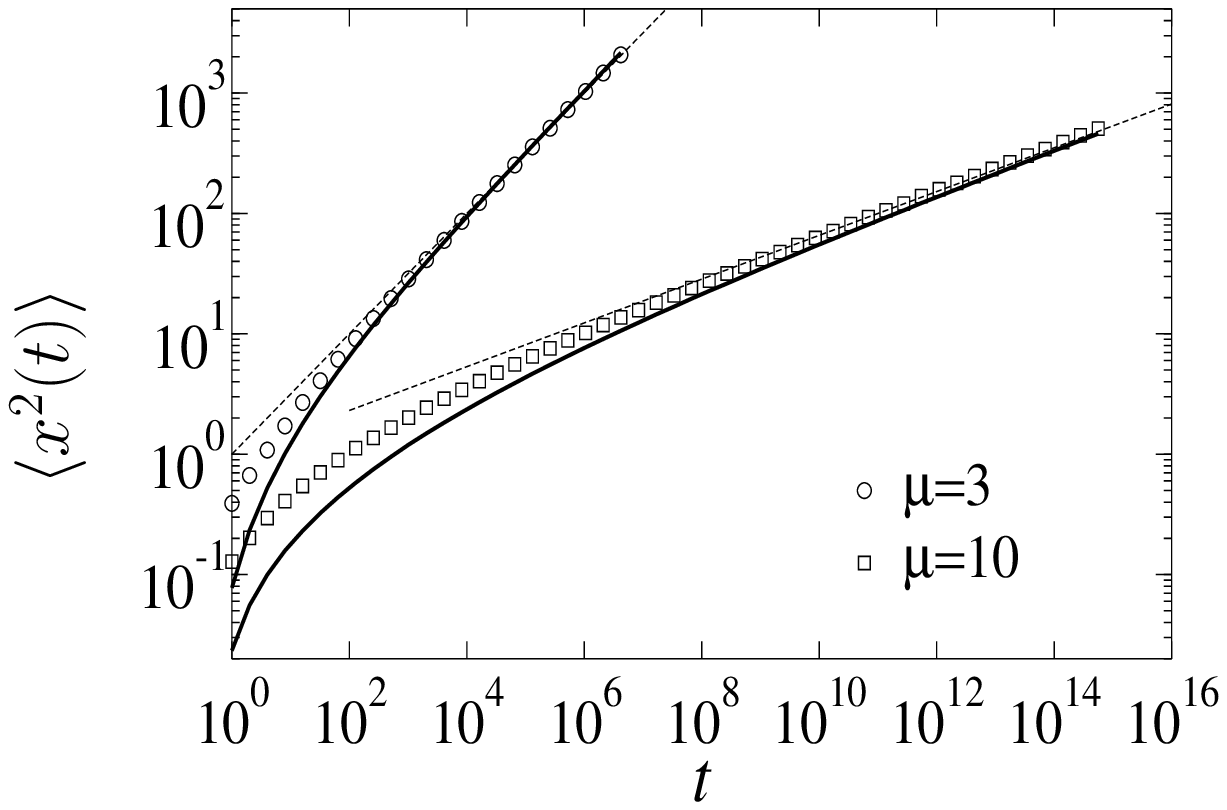,width=8cm}\par
\epsfig{file=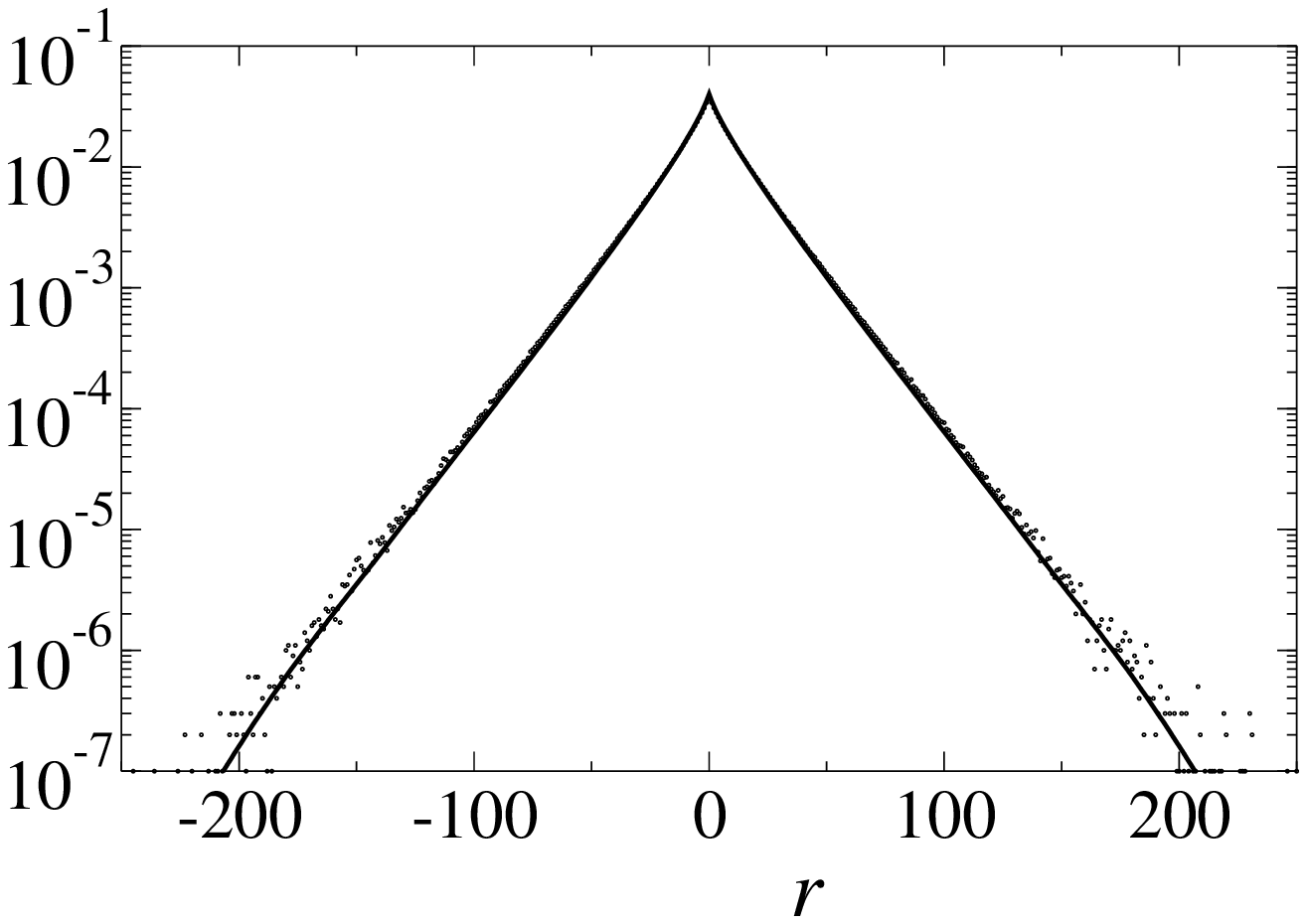,width=8cm}
\epsfig{file=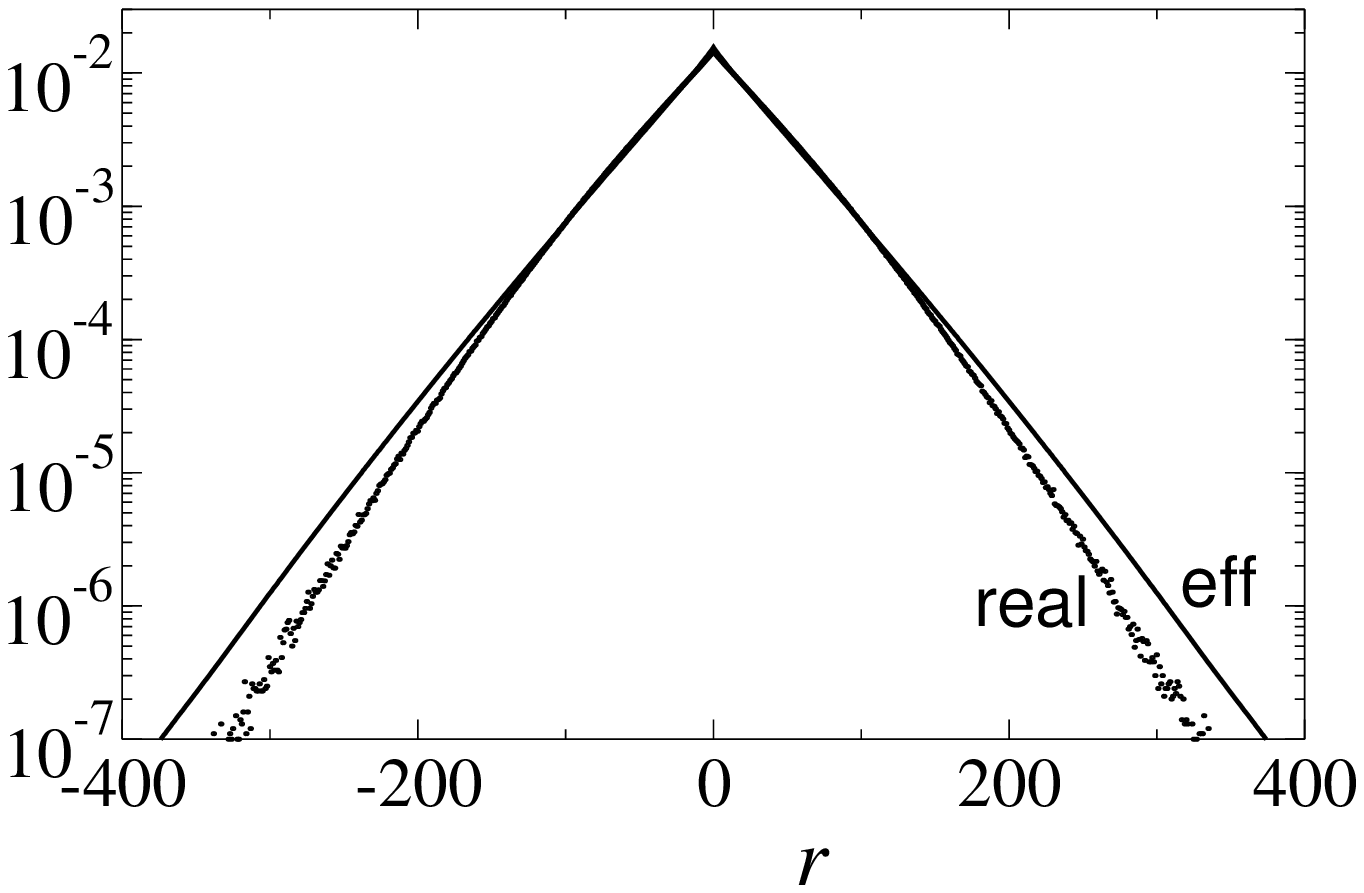,width=8cm}\par
\caption{
(Top) Plots of $\langle x^2(t) \rangle$, comparing real
dynamics (symbols)
and effective dynamics (solid lines).  
The dashed lines show the power
law scaling of (\ref{equ:scaling}), with unit prefactors
(we plot $\langle x^2\rangle=t^{2/(1+\mu)}$).
The effective dynamics
captures the time-dependence of the mean-squared displacement
quantitatively, even at the relatively 
small value $\mu=3$.  (Bottom) We show the disorder-averaged
diffusion front $\overline{G}_n(t)$ for real dynamics (points)
and effective dynamics (lines) at $\mu=10$ (left) and $\mu=3$ (right).  The times
are those of the final points in the top panel, $t=2^{49}$
and $t=2^{22}$: the data of that panel indicate that
these times are long enough to be representative of the
asymptotic large-time scaling.  The effective dynamics agrees well
with the real dynamics for $\mu=10$, but for the smaller $\mu$ the agreement
deteriorates because the assumption of well-separated time 
scales breaks down.
} \label{fig:compare_eff}
\end{figure}

In Fig.~\ref{fig:compare_eff}, 
we compare the predictions of the effective dynamics
with simulations of the trap and barrier models.  At
long times, we find 
good quantitative agreement at $\mu=10$.  At $\mu=3$, the effective
dynamics still captures the width of the distribution, although 
the absence of well-separated time scales leads to deviations
in the tail.  The diffusion front for the trap model
was considered 
in~\cite{Bertin03}, where it was predicted to decay as
$\overline{G}_n(t) \sim |n|^{(-\mu+1)/2\mu} \exp[ -b |n|^{(\mu+1)/\mu} ]$,
for large $n$ and $\mu$ close to unity (in our notation, the parameter
$\mu$ of~\cite{Bertin03} is denoted by $1/\mu$).
While our results are consistent with this prediction, 
our statistics are not good enough to accurately determine the asymptotic
scaling of the tails of the diffusion front.
Overall, we would argue that our results indicate that the proposed
effective dynamics does indeed capture the physical processes
responsible for propagation in the trap and barrier models.

\section{Conclusions}
\label{sec:conc}

We have discussed several implications of the duality between
one-dimensional trap and barrier models, including~(\ref{equ:local_dual}),
which is an exact local relation between their diffusion fronts valid
for fixed disorder, and (\ref{equ:ave_dual}), which states
that their disorder-averaged propagators are equal.
In addition, we have presented a unified effective dynamics for diffusion
fronts in the large $\mu$ limit
of the trap and barrier models.  This dynamics
captures the scaling of the diffusion front, and gives
good quantitative estimates of its width.  
Our arguments indicate that it gives the exact shape of the diffusion
front in the limit of extreme subdiffusion ($\mu\to\infty$),
with agreement becoming less good as the system approaches
the diffusive regime ($\mu<1$).

The effective dynamics exploits the fact that in both
models, the time scale on which traps and barriers become 
irrelevant depend on both their rates $W$, and on the spacings 
between them.  In the barrier model, the dependence on spacings
arises from the fact that
if the relevant barriers are well-separated, the particle spends only a
small fraction of its time adjacent to them.
In the trap model, it originates from the fact that a particle leaving
a given trap is likely to be reabsorbed there before it can reach
a neighbouring deep trap. Despite the different physical mechanisms,
these two processes affect the diffusion front in closely related
ways, and mathematically result in the master operators for these processes
having equal eigenspectra.

Thus, for the purposes of the diffusion front, it is indeed appropriate
to consider the barrier model as representing an environment of effective
traps in which a particle diffuses.  However, as we discussed, the
trajectories by 
which the system evolves in each case are different. The effective
dynamics illustrates the nature of these trajectories: in the barrier
model, the particle diffuses rapidly within its effective trap, leading
to a number of hops growing as $N\sim t$.  In the trap model, the particle
is localised in deep traps, making excursions from its current trap which
mostly lead to reabsorption within that trap.  The mean hopping rate
decreases with time as the system ages~\cite{MonBou96}, 
leading to $N\sim t^{2/z}$.  It would be
illuminating to investigate whether the duality
relation~(\ref{equ:local_dual}) at fixed
disorder can be cast as a relation between the particle trajectories
in the barrier and trap models,
to see whether these different scalings emerge naturally.

Given the different physical processes underlying the dynamics
of the models, we argue that the relationships between the 
diffusion fronts are, to a certain extent, coincidental.  They do not
represent a physical equivalence of the models themselves,
and they do not generalise in a simple way to 
correlation functions other than the diffusion front.  
However, it seems to us that
the equality of the disorder-averaged propagators 
and the associated algebraic structure might be
exploited in further analytic work on these models.

\ack
We thank Peter Mayer for very helpful discussions.  
RLJ was funded initially by NSF grant CHE-0543158
and later by Office of Naval Research Grant 
No.~N00014-07-1-0689.


\begin{thebibliography}{99}
\bibitem{subdiff-note}
Subdiffusive propagation is characterised by a dynamical exponent
$z$, which describes the relative scaling of space and time
at a critical point.  For example,
the fragile models considered in~\cite{subdiff-scaling}
exhibit subdiffusive behaviour, with $z>2$.  In the Fredrickson-Andersen
model considered in~\cite{FA-scaling}, the scaling at the critical
point is diffusive, with $z=2$.  Away from criticality, the 
correlation length $\xi$ and correlation time $\tau$ satisfy 
$D\tau\sim\xi^2$, where $D$ is the diffusion constant.  However, 
$D$ itself has a non-trivial temperature dependence, so the
temperature scaling cannot in general be written as $\tau\sim\xi^2$.
\bibitem{parity-proof}
To see this, use (\ref{equ:w_trap}) to obtain
$(\mathrm{d}/\mathrm{d}t) \sum_{n=N}^{M} 
(x_n-x_k) \Gt_{n\leftarrow k}(t)=(x_{M}-x_{N})
[W_{N}\Gt_{N\leftarrow k}(t) - W_{M} \Gt_{M\leftarrow k}(t)]$.  
Taking site $M$
far to the right of site $k$, and site $N$ far to its left,
this time derivative vanishes.  Therefore, using the initial condition
for $\Gt$, we have
$\sum_n(x_n-x_k)\Gt_{n\leftarrow k}(t)=0$.  Hence,
$\langle [x(t)-x(t')]x(t') \rangle =
\sum_{nk} (x_n-x_k)x_k\Gt_{n\leftarrow k}(t-t')\Gt_{n\leftarrow i}(t')=0$.
\end{thebibliography}
\end{document}